\documentclass{webofc}\usepackage[varg]{txfonts} 
\usepackage{bm}\usepackage{color}\usepackage{booktabs}
\woctitle{QCD@Work 2018}\begin{document}
\title{Exotic Tetraquark Mesons in Large-$\bm{N_{\rm c}}$ Limit:\\
an Unexpected Great Surprise}\author{Wolfgang Lucha\inst{1}
\fnsep\thanks{\email{Wolfgang.Lucha@oeaw.ac.at}}\and Dmitri
Melikhov\inst{1,2,3}\fnsep\thanks{\email{dmitri_melikhov@gmx.de}}
\and Hagop Sazdjian\inst{4}\fnsep
\thanks{\email{sazdjian@ipno.in2p3.fr}}}\institute{Institute for
High Energy Physics, Austrian Academy of Sciences,
Nikolsdorfergasse 18,\\A-1050 Vienna, Austria\and D.~V.~Skobeltsyn
Institute of Nuclear Physics, M.~V.~Lomonosov Moscow State
University,\\119991 Moscow, Russia\and Faculty of Physics,
University of Vienna, Boltzmanngasse 5, A-1090 Vienna, Austria\and
Institut de Physique Nucl\'eaire, CNRS-IN2P3, Universit\'e
Paris-Sud, Universit\'e Paris-Saclay,\\91405 Orsay Cedex, France}

\abstract{Two-ordinary-meson scattering in large-$N_{\rm c}$ QCD
implies consistency criteria for intermediate-tetraquark
contributions. Their fulfilment at $N_{\rm c}$-leading order
constrains the nature of the spectrum of genuinely exotic
tetraquark states.}\maketitle

\section{Introduction: $\bm{N_{\rm c}}$-Leading First-Principles
Approach to Tetraquarks}Qualitative information on systems
controlled by quantum chromodynamics may be collected by
considering large-$N_{\rm c}$ QCD \cite{GH,EW}, a quantum field
theory generalizing QCD by enabling the number $N_{\rm c}$ of
colour degrees of freedom to differ from $N_{\rm c}=3$ and grow
beyond bounds~while, simultaneously, demanding that the product
$N_{\rm c}\,\alpha_{\rm s}$ of $N_{\rm c}$ and the strong
fine-structure coupling $$\alpha_{\rm s}\equiv\frac{g_{\rm
s}^2}{4\pi}$$approaches a finite value in the large-$N_{\rm c}$
limit. For the $N_{\rm c}$ behaviour of $\alpha_{\rm s},$ this
demand~implies$$\alpha_{\rm s}\propto\frac{1}{N_{\rm
c}}\qquad\mbox{for}\quad N_{\rm c}\to\infty\ .$$In an attempt to
stay as close as possible to intuition, we adhere to the
presumably very natural (but clearly not compulsory) assumption
that the fermionic dynamical degrees of freedom,~the quarks,
continue to transform according to the $N_{\rm c}$-dimensional,
fundamental~representation~of the gauge group ${\rm SU}(N_{\rm
c}).$ By utilizing QCD's large-$N_{\rm c}$ limit ($N_{\rm
c}\to\infty$) and $1/N_{\rm c}$ expansion~(in powers of $1/N_{\rm
c}$) about this limit, we extract constraints on crucial features
(\emph{e.g.}, decay~widths) \cite{TQ1,TQ4} of tetraquarks, meson
bound states of two quarks and two antiquarks predicted by QCD.

With respect to their \emph{flavour\/} degrees of freedom,
tetraquarks can be classified (Table~\ref{C})~by specifying ---
for the two quarks and two antiquarks constituting the tetraquark
bound state~---\begin{itemize}\item the number of different quark
flavours encountered in such bound state, in combination~with\item
the total number of \emph{open\/} quark flavours, defined as the
number of quark flavours that are~not counterbalanced by an
antiquark of same flavour and hence carried by the observed
mesons.\end{itemize}\pagebreak Whereas, by definition, for
ordinary mesons a count of open flavour should yield zero or two,
the case of four open flavours signals the tetraquark nature of
the respective quark bound~state.

\begin{table}[t]\begin{center}\caption{Classification of
tetraquark mesons by content of \emph{different\/} and
\emph{open\/} quark flavour. The notion \emph{open-flavour
number\/} relates to the net sum of flavours not compensated by a
corresponding antiflavour.}\label{C}\begin{tabular}{ccc}\toprule
number of different&$\quad$ \ tentative quark configuration \
$\quad$&number of open\\quark flavours involved&$\bar
q_\square\,q_\square\ \,\bar q_\square\,q_\square$&quark flavours
involved\\\midrule4&$\bar q_1\,q_2\ \,\bar
q_3\,q_4$&4\\[.23ex]\hline\\[-2.1ex]3&$\bar q_1\,q_2\ \,\bar
q_3\,q_2$&4\\&$\bar q_1\,q_2\ \,\bar q_1\,q_3$&4\\&$\bar q_1\,q_2\
\,\bar q_2\,q_3$&2\\&$\bar q_1\,q_2\ \,\bar
q_3\,q_3$&2\\[.23ex]\hline\\[-2.1ex]2&$\bar q_1\,q_2\ \,\bar
q_1\,q_2$&4\\&$\bar q_1\,q_1\ \,\bar q_1\,q_2$&2\\&$\bar q_1\,q_1\
\,\bar q_2\,q_1$&2\\&$\bar q_1\,q_2\ \,\bar q_2\,q_1$&0\\&$\bar
q_1\,q_1\ \,\bar q_2\,q_2$&0\\[.23ex]\hline
\\[-2.1ex]1&$\bar q_1\,q_1\ \,\bar q_1\,q_1$&0\\\bottomrule
\end{tabular}\end{center}\end{table}

We focus to two variants of tetraquarks with particularly
interesting quark flavour content:\begin{enumerate}\item
\emph{flavour-exotic\/} tetraquarks $T=(\bar q_1\,q_2\,\bar
q_3\,q_4)$ with all four (anti-) quark flavours different;\item
\emph{flavour-cryptoexotic\/} tetraquarks $T=(\bar q_1\,q_2\,\bar
q_2\,q_3)$ involving three different (anti-) quark flavours, by
containing a quark--antiquark pair of same flavour differing from
the others.\end{enumerate}

In order to work out --- at least, at a qualitative level --- some
basic features of such~kind of tetraquarks, we investigate, for
two ordinary mesons of appropriate flavour quantum numbers, their
possible scattering reactions into two ordinary mesons with regard
to potential $s$-channel contributions of intermediate poles
interpretable as a manifestation of tetraquarks with narrow decay
width. For both sets of tetraquark in our focus of interest, as
well as for the one with two different flavours but no open
flavour, we have to analyze two variants of scattering~processes:
\begin{itemize}\item flavour-preserving ones, with identical
flavour content of the initial- and final-state mesons;\item
flavour-rearranging ones, with unequal flavour content of the
initial- and final-state mesons.\end{itemize}We identify all the
contributions to four-point correlation functions of quark
bilinear operators $j_{ij}\equiv\bar q_i\,q_j$ interpolating
ordinary mesons $M_{ij}$ (notationally exploiting the actual
irrelevance of parity and spin therein) that are capable of
supporting a pole related to a tetraquark built up by four (anti-)
quarks of masses $m_i$, $i=1,\dots,4,$ by imposing an unambiguous
selection criterion \cite{TQ1,TQ4}: for the scattering of two
mesons of momenta $p_1$ and $p_2,$ an allowable Feynman diagram
not only has to depend on the Mandelstam variable
$s\equiv(p_1+p_2)^2$ in a non-polynomial way but has to admit a
four-quark intermediate state with related branch cut starting at
the branch~point$$s=(m_1+m_2+m_3+m_4)^2\ .$$

\section{Four different quark flavours
\boldmath{\,$\Longleftrightarrow$\,} flavour-exotic tetraquark
meson}
\begin{figure}[t]\centering
\includegraphics[scale=.43668,clip]{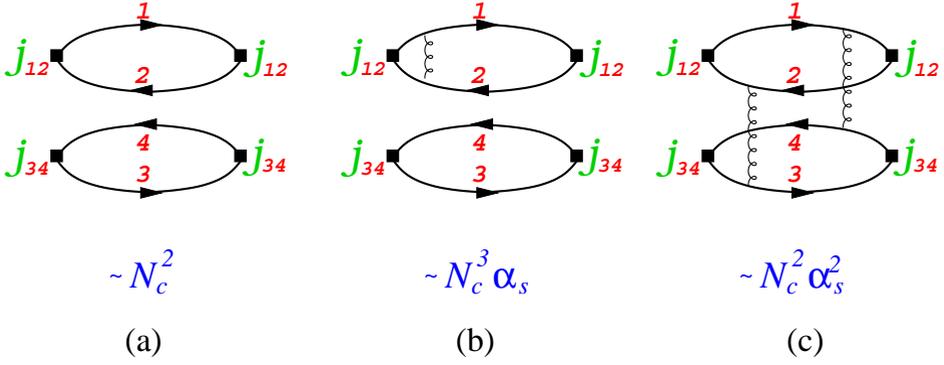}
\caption{Flavour-preserving four-point correlation function of
quark bilinear operators $j_{ij}$: examples~of Feynman diagrams
contributing at low order to this correlator's $1/N_{\rm c}$
expansion, \emph{viz.}, at order $N_{\rm c}^2$ (a,b) or $N_{\rm
c}^0$ (c) \cite[Fig.~1]{TQ1}. Tetraquark-friendly contributions of
lowest order in $1/N_{\rm c}$ turn out to be of order~$N_{\rm
c}^0$~(c).}\label{fe1}\end{figure}

\begin{figure}[t]\centering
\includegraphics[scale=.4146,clip]{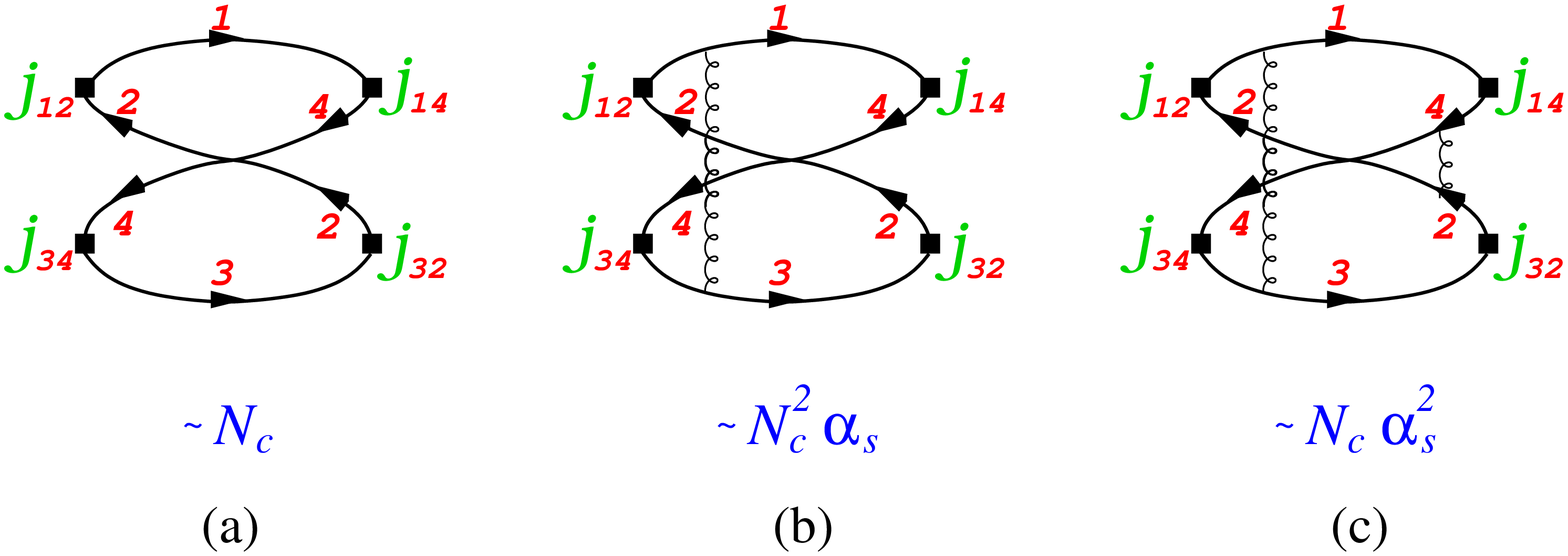}
\caption{Flavour-rearranging four-point correlation function of
quark bilinear currents $j_{ij}$: examples~of Feynman diagrams
contributing at low order to this correlator's $1/N_{\rm c}$
expansion, that is, at order $N_{\rm c}$ (a,b) or $N_{\rm c}^{-1}$
(c) \cite[Fig.~2]{TQ1}. Tetraquark-phile contributions of lowest
order in $1/N_{\rm c}$ prove to be of order $N_{\rm c}^{-1}$~(c).}
\label{fe2}\end{figure}

Surprisingly or not, for truly flavour-exotic tetraquark
states, $T=(\bar q_1\,q_2\,\bar q_3\,q_4),$ the leading-$N_{\rm
c}$ dependence of all contributions to the correlation functions
of four quark bilinear operators $j_{ij}$ (exemplified in
Figs.~\ref{fe1} and \ref{fe2}) potentially capable of developing
this tetraquark pole \emph{differs\/}, for the flavour-preserving
(Fig.~\ref{fe1}(c)) and flavour-rearranging (Fig.~\ref{fe2}(c))
cases, by one order~of~$N_{\rm c}$:$$\langle
j^\dag_{12}\,j^\dag_{34}\,j_{12}\,j_{34}\rangle_{\rm T}=O(N_{\rm
c}^0)\ ,\qquad\langle
j^\dag_{14}\,j^\dag_{32}\,j_{14}\,j_{32}\rangle_{\rm T}=O(N_{\rm
c}^0)\ ,\qquad\langle
j^\dag_{14}\,j^\dag_{32}\,j_{12}\,j_{34}\rangle_{\rm T}=O(N_{\rm
c}^{-1})\ .$$



\noindent The resulting $N_{\rm c}$-leading
tetraquark--two-ordinary-meson amplitudes $A$ imply the
presence~of, at least, two\footnote{In such a situation, their
conclusions some people in the form may phrase ``always two there
are, \dots, no less''~\cite{Y}.} tetraquarks $T_{A,B}$ with,
however, decay rates $\Gamma(T_{A,B})$ of similar large-$N_{\rm
c}$ decrease:\begin{align*}\underbrace{A(T_A\longleftrightarrow
M_{12}\,M_{34})=O(N_{\rm c}^{-1})}
_{\mbox{$\Longrightarrow\qquad\Gamma(T_A)=O(N_{\rm c}^{-2})$}}
\qquad\stackrel{N_{\rm c}}{>}\qquad A(T_A\longleftrightarrow
M_{14}\,M_{32})=O(N_{\rm c}^{-2})\ ,&\\[1ex]
A(T_B\longleftrightarrow M_{12}\,M_{34})=O(N_{\rm
c}^{-2})\qquad\stackrel{N_{\rm c}}{<}\qquad
\underbrace{A(T_B\longleftrightarrow M_{14}\,M_{32})=O(N_{\rm
c}^{-1})}_{\mbox{$\Longrightarrow\qquad\Gamma(T_B)=O(N_{\rm
c}^{-2})$}}\ .&\end{align*}

\section{Two open quark flavours \boldmath{\,$\equiv$\,}
flavour-cryptoexotic tetraquark meson}
\begin{figure}[h]\centering
\includegraphics[scale=.43668,clip]{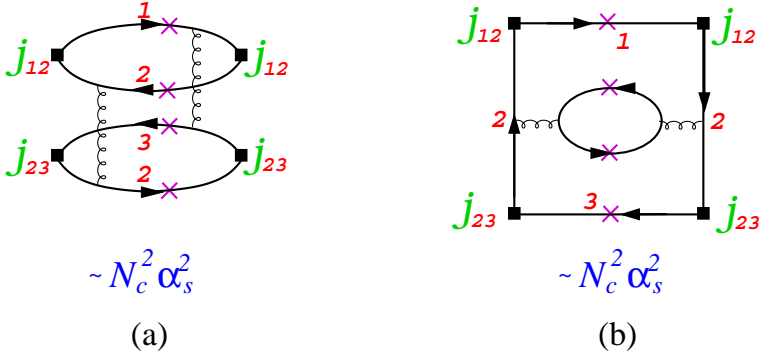}
\caption{Flavour-preserving four-point correlation function of
quark bilinear operators $j_{ij}$: examples of tetraquark-phile
Feynman diagrams contributing at the $N_{\rm c}$-leading order
$N_{\rm c}^0$ to this correlation function's $1/N_{\rm c}$
expansion \cite[Fig.~3]{TQ1}. (Purple crosses indicate quarks
potentially contributing to a tetraquark pole.)}\label{ce1}
\end{figure}

\begin{figure}[h]\centering
\includegraphics[scale=.43668,clip]{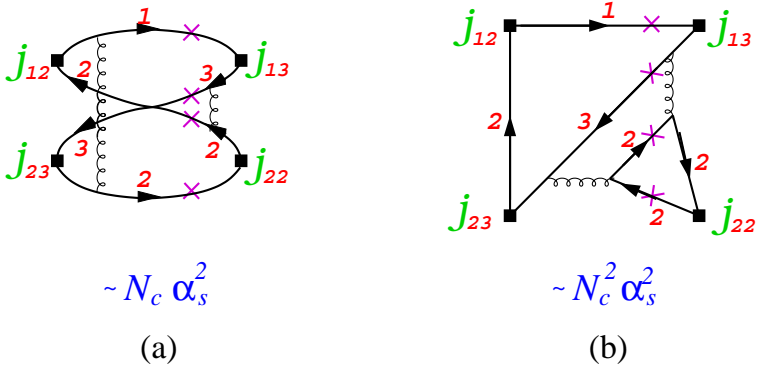}
\caption{Flavour-rearranging four-point correlation function of
quark bilinear currents $j_{ij}$: examples~of tetraquark-phile
Feynman diagrams contributing at low order to such correlator's
$1/N_{\rm c}$ expansion, \emph{i.e.},~at order $N_{\rm c}^{-1}$
(a) or $N_{\rm c}^0$ (b) \cite[Fig.~4]{TQ1}; among these, the
$N_{\rm c}$-leading contributions prove to be of order $N_{\rm
c}^0$~(b).}\label{ce2}\end{figure}

In the case of
flavour-cryptoexotic tetraquark mesons built from three different
quark flavours, $T=(\bar q_1\,q_2\,\bar q_2\,q_3),$ the
large-$N_{\rm c}$ behaviour of flavour-preserving (Fig.~\ref{ce1})
and flavour-reshuffling (Fig.~\ref{ce2}) subcategories of those
contributions to the correlation functions of four quark bilinear
currents $j_{ij}$ which might support the development of a
tetraquark pole turns out to be identical:$$\langle
j^\dag_{12}\,j^\dag_{23}\,j_{12}\,j_{23}\rangle_{\rm T}=O(N_{\rm
c}^0)\ ,\qquad\langle
j^\dag_{13}\,j^\dag_{22}\,j_{13}\,j_{22}\rangle_{\rm T}=O(N_{\rm
c}^0)\ ,\qquad\langle
j^\dag_{13}\,j^\dag_{22}\,j_{12}\,j_{23}\rangle_{\rm T}=O(N_{\rm
c}^0)\ .$$



\noindent So, a single tetraquark $T_{\rm C}$ (which, due to its
cryptoexotic nature, can mix with the meson $M_{13}$) satisfies
all constraints induced by the $N_{\rm c}$-leading
tetraquark--two-ordinary-meson amplitudes$$\underbrace{A(T_{\rm
C}\longleftrightarrow M_{12}\,M_{23})=O(N_{\rm
c}^{-1})\qquad\stackrel{N_{\rm c}}{=}\qquad A(T_{\rm
C}\longleftrightarrow M_{13}\,M_{22})=O(N_{\rm c}^{-1})}
_{\mbox{$\Longrightarrow\qquad\Gamma(T_{\rm C})=O(N_{\rm
c}^{-2})$}}\ .$$

\section{Insights: $\bm{N_{\rm c}}$-Leading Conclusions for
(Crypto-) Exotic Tetraquarks}In summary, we find that
self-consistency conditions arising from the inspection of
tetraquark contributions to the scattering amplitudes of two
ordinary mesons into two ordinary mesons in the $1/N_{\rm c}$
expansion of large-$N_{\rm c}$ QCD provide rigorous constraints on
the features of tetraquark states \cite{TQ1,TQ4}. Demanding these
constraints to be satisfied at $N_{\rm c}$-leading order implies
\cite{TQ1,TQ4}~that\begin{itemize}\item genuinely exotic
tetraquarks need to appear in pairs $(T_A,T_B)$ the members of
which differ in the large-$N_{\rm c}$ behaviour of their dominant
or preferred decay modes to two ordinary mesons;\item both
genuinely exotic ($T_{A,B}$) and cryptoexotic ($T_C$) tetraquarks
exhibit narrow decay widths$$\Gamma(T)\propto1/N_{\rm
c}^2\xrightarrow[N_{\rm c}\to\infty]{}0\qquad\mbox{for}\quad
T=T_A,T_B,T_{\rm C}\ .$$\end{itemize}Table \ref{W} confronts these
findings for the rates of the large-$N_{\rm c}$ decrease of the
total decay widths of exotic and cryptoexotic tetraquarks with
corresponding outcomes of earlier analyses \cite{KP,CL,MPR}.
Imposition of additional requirements clearly may strengthen the
predicted large-$N_{\rm c}$ decrease. Differences to our results
arise from misidentifying the actually $N_{\rm c}$-leading
contribution to the tetraquark pole or from consideration of
merely a single (say, the flavour-reshuffling) channel.

\begin{table}[h]\begin{center}\caption{Comparison: predictions of
\emph{upper bounds\/} on the large-$N_{\rm c}$ behaviour of
tetraquark decay rates.}\label{W}\begin{tabular}{lccr}\toprule
Author Collective \ $\quad$&\multicolumn{2}{c}{Decay Width
$\Gamma$}&$\quad$ \ Reference\\[.1ex]&Exotic Tetraquarks&
Cryptoexotic Tetraquarks&\\\midrule Lucha \emph{et al.}&
$O(1/N_{\rm c}^2)$&$O(1/N_{\rm c}^2)$&\cite{TQ1,TQ4}\\[.1ex]Knecht
and Peris&$O(1/N_{\rm c}^2)$&$O(1/N_{\rm c})$&
\cite{KP}\\[.1ex]Cohen and Lebed&$O(1/N_{\rm c}^2)$&---&\cite{CL}
\\[.1ex]Maiani \emph{et al.}&$O(1/N_{\rm c}^3)$&$O(1/N_{\rm c}^3)$
&\cite{MPR}\\\bottomrule\end{tabular}\end{center}\end{table}

\section*{Acknowledgements}D.~M.\ is grateful for support by the
Austrian Science Fund (FWF) under project P29028-N27.

\end{document}